\documentclass[sigplan,screen]{acmart}
\AtBeginDocument{%
  }

\copyrightyear{2018}
\acmYear{2018}
\acmDOI{XXXXXXX.XXXXXXX}
\acmConference[GPGPU 2025]{General Purpose Processing on Graphics Processing Units}{Mar 1, 2025}{Las Vegas, NV}

\acmISBN{978-1-4503-XXXX-X/18/06}

\renewcommand\footnotetextcopyrightpermission[1]{} 
\usepackage{lipsum}
\usepackage{tikz} 
\usepackage{csquotes}
\usepackage{multirow}
\usepackage{hyperref} 
\usepackage{todonotes}
\newcommand{\tsc}[1]{\textsuperscript{#1}}
\usepackage{enumitem}
\setlist[itemize]{leftmargin=3.0mm}
\renewcommand{\lipsum}[1][]{} 

\begin{document}

\title{Can Tensor Cores Benefit Memory-Bound Kernels? (No!)}

\author{Lingqi Zhang\tsc{1}, Jiajun Huang\tsc{2}, Sheng Di\tsc{3}, Satoshi Matsuoka\tsc{1}, Mohamed Wahib\tsc{1}}
\affiliation{
  \institution{\tsc{1} RIKEN Center for Computational Science, Japan, {\small\{lingqi.zhang@riken.jp, matsu@acm.org, mohamed.attia@riken.jp\}}}
  \institution{\tsc{2} University of California, Riverside, USA, {\small\{jhuan380@ucr.edu\}}}
  \institution{\tsc{3} Argonne National Laboratory, USA, {\small\{sdi1@anl.gov\}}}
  \country{}
}

\renewcommand{\shortauthors}{Lingqi Z. et al.}
\renewcommand{\authors}{Lingqi Zhang, Jiajun Huang, Sheng Di, Satoshi Matsuoka and Mohamed Wahib}




\begin{abstract}
Tensor cores are specialized processing units within GPUs that have demonstrated significant efficiency gains in compute-bound applications such as Deep Learning Training by accelerating dense matrix operations. Given their success, researchers have attempted to extend tensor core capabilities beyond dense matrix computations to other computational patterns, including memory-bound kernels. Recent studies have reported that tensor cores can outperform traditional CUDA cores even on memory-bound kernels, where the primary performance bottleneck is not computation.
In this research, we challenge these findings through both theoretical and empirical analysis. Our theoretical analysis reveals that tensor cores can achieve a maximum speedup of only $1.33$× over CUDA cores for memory-bound kernels in double precision (for V100, A100, and H100 GPUs). We validate this theoretical limit through empirical analysis of three representative memory-bound kernels-STREAM Scale, SpMV, and stencil. We demonstrate that optimizing memory-bound kernels using tensor cores does not yield sound performance improvements over CUDA cores.

\end{abstract}

 \maketitle

\section{Introduction}
Since Nvidia introduced tensor cores in their Volta architecture in 2017~\cite{8344474}, researchers have extensively explored their applications across various domains, including dense linear algebra~\cite{bhaskaracharya2020automatickernelgenerationvolta}, sparse linear algebra~\cite{10.1145/3620666.3651378,okanovic2024high}, and spectral methods~\cite{9563043,9460474}. The low-precision computation capabilities of tensor cores, which offer substantially higher performance, have spawned innovative approaches such as mixed-precision algorithms~\cite{haidar2018harnessing,lu2024amgt} and precision recovery~\cite{10.1145/3650200.3656634,ootomo2022recovering,ootomo2024dgemm}. However, despite this broad adoption, fundamental analysis of tensor core performance remains limited, with only a few studies focusing on microbenchmarking~\cite{9931992,10579250,9926299}.


While tensor cores represent a powerful tool for program acceleration, their effective utilization requires a thorough understanding of their performance characteristics. This research addresses this knowledge gap by focusing specifically on memory-bound kernels, which constitute a significant portion of HPC workloads~\cite{austinsystem}. Our study seeks to answer: 



\begin{itemize}
\item What is the theoretical performance ceiling for tensor cores when applied to memory-bound kernels?
\item Do current tensor core implementation strategies provide real performance benefits for memory-bound kernels?

\end{itemize}
To address these questions, we make the following contributions:
\begin{itemize}
\item A comprehensive theoretical analysis of tensor core performance for memory-bound kernels.
\item An empirical evaluation comparing tensor core implementations against their CUDA core counterparts across representative memory-bound kernels.
\end{itemize}


The rest of the paper is organized as follows. Section~\ref{sec:background} provides essential background concepts. Section~\ref{sec:workload} introduces our studied memory-bound workloads. Section~\ref{sec:theory} analyzes two extreme scenarios to establish tensor core performance bounds. Section~\ref{sec:eval} empirically substantiates our claims using representative memory-bound kernels. Finally, we present key takeaways and conclusions.


\section{Background}\label{sec:background}

\subsection{Tensor Core}\label{sec:tensorcore}

\begin{figure}[t!]
    \centering
    \includegraphics[width=\linewidth]{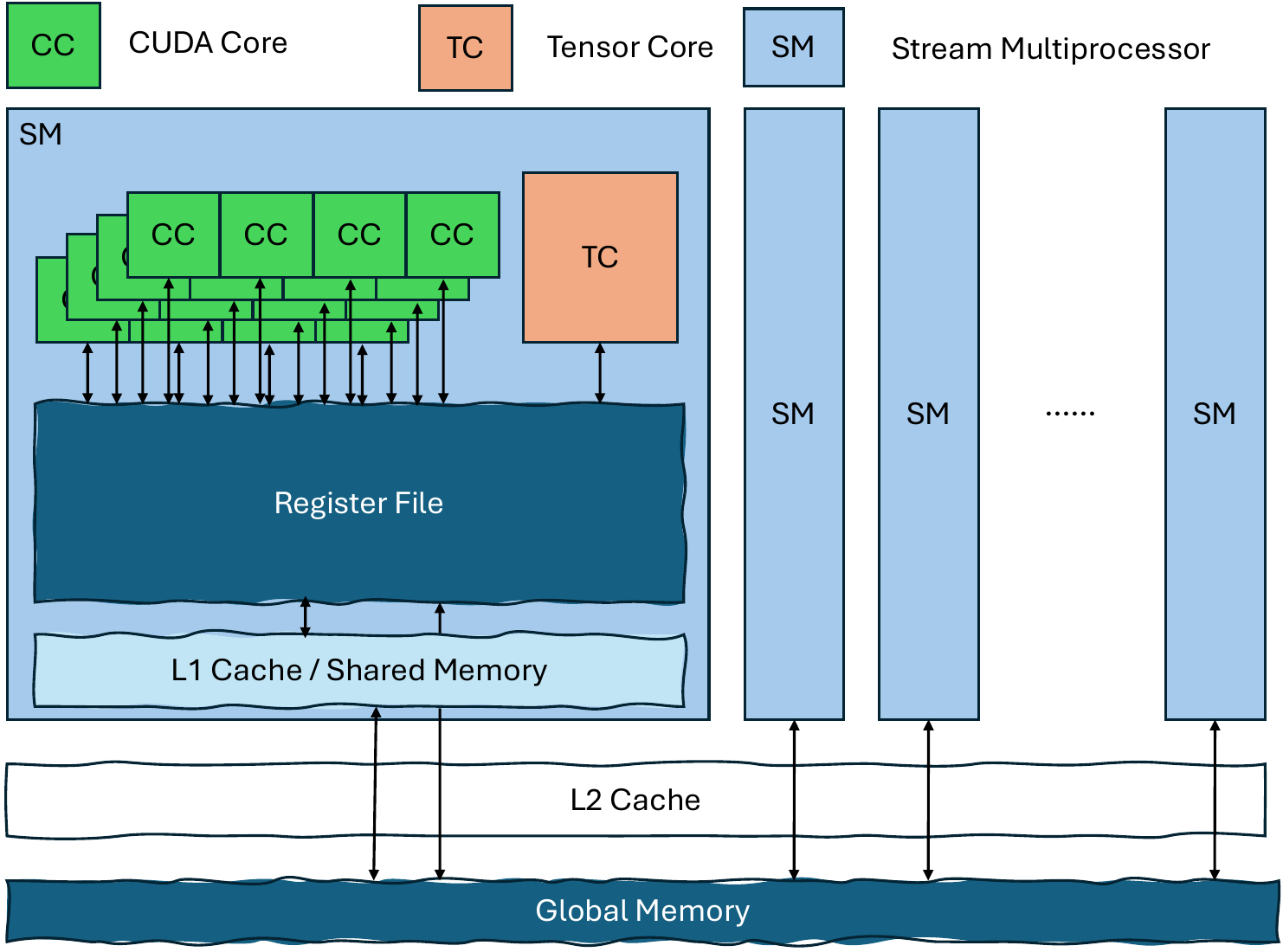}
    \vspace{-25pt}
    \caption{Nvidia GPU memory hierarchy.}
    \label{fig:tcarch}
\end{figure}

Tensor Core is a specialized systolic array matrix engine~\cite{9460517} integrated within the Stream Multiprocessor (SM) of modern Nvidia GPUs. It operates alongside traditional CUDA cores. The execution pathway for both units follows the GPU's memory hierarchy: data is first loaded from global memory into the register file, from which either CUDA cores or Tensor Cores can access it for computation. This memory access abstraction aligns with previous studies~\cite{10579250,9931992}. Figure~\ref{fig:tcarch} illustrates this memory hierarchy and the relationship between different memory levels in Nvidia GPUs.



\subsection{Machine Balance} 

We define machine balance ($\mathbb{B}$)~\cite{mccalpin1995memory} as the ratio between peak computational performance ($P$) and memory bandwidth ($B$):
\begin{equation}\footnotesize
\mathbb{B}=\tfrac{P}{B}
\end{equation}
\subsection{Roofline Model}
The roofline model~\cite{williams2009roofline,ofenbeck2014applying} provides an upper-bound performance prediction framework based on operational intensity ($\mathbb{I}$), which is defined as the ratio of computational work ($\mathbb{W}$) to memory traffic ($\mathbb{Q}$):
\begin{equation}\footnotesize
\mathbb{I}=\tfrac{\mathbb{W}}{\mathbb{Q}}
\end{equation}
The model calculates attainable performance ($\mathbb{P}$) as:
\begin{equation}\footnotesize
\mathbb{P}=\min(P, B\times \mathbb{I})
\end{equation}
This visualization framework helps identify system bottlenecks and performance limits.
\subsection{Tensor Core in the Roofline Model}
As discussed in Section~\ref{sec:tensorcore}, tensor cores can be represented as an additional performance ceiling above the CUDA core baseline in the roofline model, because tensor cores and CUDA cores share the same memory hierarchy and cannot operate simultaneously due to the Dark Silicon Effect. This roofline abstraction aligns with an existing study~\cite{yang2020hierarchical}.

\subsection{Relationship Between Machine Balance and Roofline Model}
The roofline model provides a framework for understanding performance bottlenecks through the interplay of two critical metrics: operational intensity and machine balance. While operational intensity ($\mathbb{I}$) characterizes a kernel's computational density, machine balance ($\mathbb{B}$) represents the hardware's ratio of computational capability to memory bandwidth. The relationship between these metrics determines whether a kernel's performance is mainly limited by computation or memory access:
\begin{equation}\footnotesize
A\ kernel\ is =
\begin{cases}
compute-bound, & \text{if } \mathbb{I}> \mathbb{B}\\
memory-bound,  & \text{if } \mathbb{I}< \mathbb{B}
\end{cases}
\end{equation}

\begin{figure*}[t]
\centering
\includegraphics[width=\linewidth]{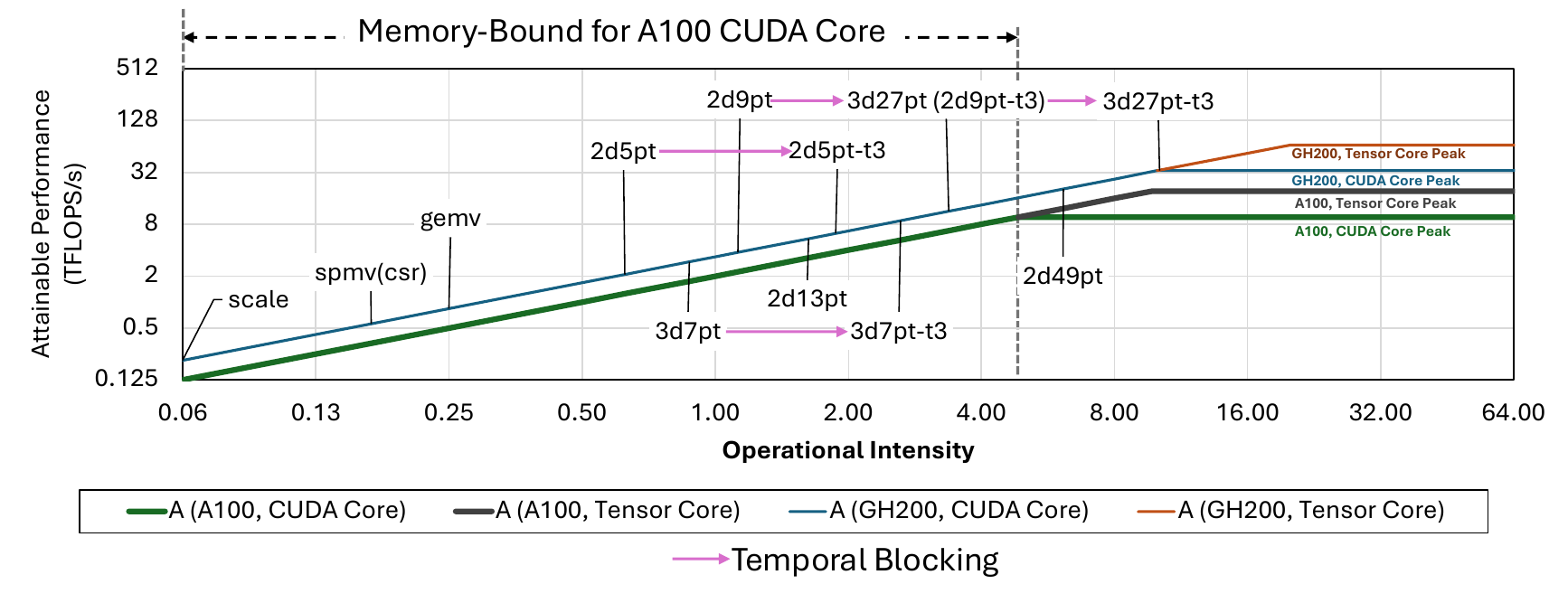}
\vspace{-30pt}
\caption{\label{fig:roofline} An example of the roofline model for both GH200 and A100-80GB GPU.
}
\end{figure*}

As illustrated in Figure~\ref{fig:roofline}, machine balance manifests as the inflection point in the roofline curve for both GH200 and A100-80GB GPUs. This point marks the transition from memory-bound to compute-bound behavior.

\section{Workloads: Memory-Bound Kernels}\label{sec:workload}

This section examines three representative memory-bound kernels: SCALE (Section~\ref{sec:theoscale}), Sparse Matrix-Vector Multiplication (SpMV) (Section~\ref{sec:theospmv}), and Stencil (Section~\ref{sec:theostencil}). We analyze these kernels through the lens of operational intensity ($\mathbb{I}$), focusing on double-precision operations (data size $\mathbb{D}=8$ bytes). 

While our analysis centers on double precision, the methodology can be extended to lower-precision scenarios.

\subsection{SCALE}\label{sec:theoscale}
SCALE, one of the STREAM benchmark~\cite{McCalpin2007}, is defined as
\begin{equation}\footnotesize
a_i = qb_i, \quad \forall i \in {1,\ldots,n}, \quad a,b \in \mathbb{R}^n, \quad q \in \mathbb{R}
\end{equation}
Each element operation requires one load, one store, and one computation, yielding: $\mathbb{W}(\text{SCALE})=1$, $\mathbb{Q}(\text{SCALE})=2\times\mathbb{D}$, and consequently $\mathbb{I}(\text{SCALE})=\tfrac{1}{16}$. STREAM benchmark is commonly used to measure sustainable memory bandwidth due to its low computational intensity.

\subsection{Sparse Matrix–Vector Multiplication (SpMV)} \label{sec:theospmv}
SpMV, crucial for iterative solvers, has format-dependent operational intensity. In this section, we begin by analyzing dense matrix-vector multiplication (GEMV) as a baseline.

\noindent\textbf{GEMV:}
For matrix $A \in \mathbb{R}^{m \times n}$ and vectors $x \in \mathbb{R}^n$, $y \in \mathbb{R}^m$, GEMV is defined as:
\begin{equation}\footnotesize
y = Ax
\end{equation}
With computing $\mathbb{W}(\text{GEMV})=m\times n\times2$ operations and memory traffic $\mathbb{Q}(\text{GEMV})=(m\times n+m+n)\times\mathbb{D}$ yields:
\begin{equation}\footnotesize
\mathbb{I}(\text{GEMV})=\frac{m\times n\times2}{(m\times n+m+n)\times\mathbb{D}}
\approx \frac{2}{\mathbb{D}}=\frac{1}{4}
\end{equation}

\noindent\textbf{SpMV:}
For sparse matrices with $nnz$ non-zeros, SpMV is defined as:
\begin{equation}\footnotesize
\begin{aligned}
y &= Ax, \
A \in \mathbb{R}^{m \times n}, \quad \text{nnz}(A) &\ll mn, \quad
x \in \mathbb{R}^n, \quad y \in \mathbb{R}^m
\end{aligned}
\end{equation}
With computation $\mathbb{W}(\text{SpMV})=2\times nnz$ and memory traffic including coordinate information $\alpha\mathbb{I}$ or packed values $\beta\mathbb{Z}$:
\begin{equation}\footnotesize
\mathbb{I}(\text{SpMV})=\frac{nnz\times2}{(nnz+m+n)\times\mathbb{D}+\alpha\mathbb{I} +\beta\mathbb{Z}}
\end{equation}
Given $nnz\ll m\times n$, we have $\mathbb{I}(\text{SpMV})<\mathbb{I}(\text{GEMV})$.

\noindent\textbf{Compressed Sparse Row (CSR) format:}
CSR format, the most common sparse representation, requires storing column indices and row pointers. With memory traffic $\mathbb{Q}(\text{SpMV,CSR})=(nnz+m+n)\times\mathbb{D}+(nnz+m+1)\times\mathbb{I}$ and computation $\mathbb{W}(\text{SpMV,CSR})=2\times nnz$:
\begin{equation}\footnotesize
\begin{split}
\mathbb{I}(\text{SpMV,CSR})&=\frac{2\times nnz}{(nnz+m+n)\times\mathbb{D}+(nnz+m+1)\times\mathbb{I}} \\
&\approx \frac{2}{\mathbb{D}+\mathbb{I}}=\frac{1}{6}<\mathbb{I}(\text{GEMV})
\end{split}
\end{equation}
This analysis confirms SpMV's memory-bound nature, consistent with prior works~\cite{10.1145/1816038.1816021,10.1145/3577193.3593705}.

\subsection{Iterative Stencils}\label{sec:theostencil}

Stencil computations is common in HPC~\cite{hagedorn2018high}. For 2D stencil, we have:
\begin{equation}\footnotesize
v(i,j) = \sum_{(p,q) \in \mathbb{S}} w_{p,q} \cdot u(i+p,j+q)
\end{equation}
where $v(i,j)$ and $u(i,j)$ are updated and original values at point $(i,j)$, and $\mathbb{S}$ defines relative offsets (e.g., 5-point stencil: $(-1,0)$, $(1,0)$, $(0,1)$, $(0,-1)$, $(0,0)$).
Ideally, only one load of $u$ and one store of $v$ are necessary:
\begin{equation}\footnotesize
\mathbb{Q}=2\times\mathbb{D}, \quad \mathbb{W}=2\times|\mathbb{S}|, \quad \mathbb{I}=\frac{|\mathbb{S}|}{\mathbb{D}}
\end{equation}
For a 2d5pt stencil where $|\mathbb{S}
(\text{2d5pt})|=5$, $\mathbb{I}(\text{2d5pt})=\tfrac{5}{8}$.

\noindent\textbf{Temporal blocking~\cite{10.1145/3577193.3593716,10.1145/3368826.3377904}} combines $t$ timesteps together:
\begin{equation}\footnotesize
\mathbb{W}_{t}=t\times2\times|\mathbb{S}|, \quad \mathbb{I}_{t}=t\times\frac{|\mathbb{S}|}{\mathbb{D}}
\end{equation}
While temporal blocking can theoretically transform memory-bound stencils into compute-bound kernels by increasing operational intensity, practical limitations exist. 

For a 2d5pt stencil on GH200 ($\mathbb{B}_{GH200}=9.99$), compute-bound behavior requires:
\begin{equation}\footnotesize
t\times \mathbb{I}(\text{2d5pt})>\mathbb{B}_{GH200} \implies t\times 0.625 > 9.99 \implies t > 15.98
\end{equation}

However, deep temporal blocking (e.g., $t > 16$) usually faces hardware limits from register pressure~\cite{10.1145/3577193.3593716,10.1145/3368826.3377904}. 

Thus, shallow temporal blocking ($t < 16$) 2d5pt stencil remains memory-bound, while deep temporal blocking might make stencil kernel register-bound.






\section{Theoretical Analysis}\label{sec:theory}
Based on the operational intensity analysis from Section~\ref{sec:workload}, Figure~\ref{fig:roofline} shows all studied kernels are memory-bound on GH200, and most memory-bound on A100. To simplify the following discussion, we assume that all kernels are throughput-bound. 



when it is throughput bound, which is usually the case in High-Performance workloads, we have time for computation $T_{\text{cmp}}=\tfrac{\mathbb{W}}{P}$. 
and time for memory access $T_{mem}=\tfrac{\mathbb{Q}}{B}$.
So we have:
\begin{equation}\footnotesize
    \tfrac{T_{mem}}{T_{cmp}}=\tfrac{\tfrac{\mathbb{Q}}{B}}{\tfrac{\mathbb{W}}{P}}=\tfrac{\mathbb{B}}{\mathbb{I}}
\end{equation}

For memory-bound kernel, $\mathbb{B}>\mathbb{I}$, we have:
\begin{equation}\footnotesize
T_{\text{mem}} > T_{\text{cmp}}
\end{equation}

We analyze two extreme cases: fully overlapped and fully un-overlapped for memory access and computation.

\subsection{Fully Overlapped}\label{sec:anaovl}
\begin{figure}[t!]
\begin{tikzpicture}[scale=0.5, every node/.style={scale=1}]
    \draw[->, thick] (0,0) -- (11,0) node[right] {\textbf\footnotesize{Time Breakdown}};
    \fill[red!70] (0,2.3) rectangle (5,3.3) node[pos=.5,black] {\footnotesize{mem}};
    \fill[blue!70] (2,1.3) rectangle (5,2.3) node[pos=.5,white] {\footnotesize{cmp(CC)}};
    \fill[yellow!70] (3,0.3) rectangle (5,1.3) node[pos=.5,black] {\footnotesize{others}};

    \draw[thick] (5,2.8) -- (6.5,2.8) node[right] {$T_{\text{mem}}$};
    \draw[thick] (5,1.8) -- (6.5,1.8) node[right] {$T_{\text{cmp}}(CC)$};
    \draw[thick] (5,0.8) -- (6.5,0.8) node[right] {$T_{\text{others}}$};

    \draw[thick] (10,1.8) -- (10,1.8) node[right] {(With CUDA Cores)};

    \fill[red!70] (0,-0.3) rectangle (5,-1.3) node[pos=.5,black] {\footnotesize{mem}};
    \fill[blue!70] (2,-1.3) rectangle (3.5,-2.3) node[pos=.5,white] {\footnotesize{cmp}};
    \fill[yellow!70] (3,-2.3) rectangle (5,-3.3) node[pos=.5,black] {\footnotesize{others}};
    
    \draw[thick] (5,-0.8) -- (6.5,-0.8) node[right] {$T_{\text{mem}}$};
    \draw[thick] (3.5,-1.8) -- (6.5,-1.8) node[right] {$T_{\text{cmp}}(TC)$};
    \draw[thick] (5,-2.8) -- (6.5,-2.8) node[right] {$T_{\text{others}}$}; 

    \draw[thick] (10,-1.8) -- (10,-1.8) node[right] {(With Tensor Cores)};
\end{tikzpicture}
\vspace{-5pt}
\caption{Fully overlapped kernel time breakdown}\label{fig:breakout_overlap}
\end{figure}
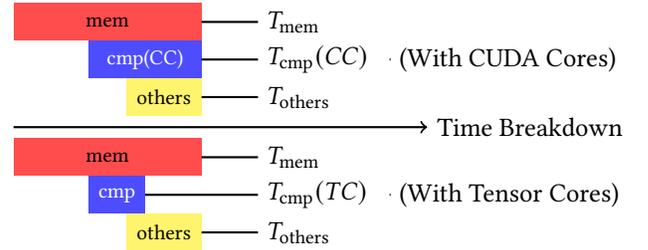

An example of an overlapped kernel time breakdown is shown in Figure~\ref{fig:breakout_overlap}. For memory-bound kernels:
\begin{equation}\footnotesize
T = \max(T_{\text{cmp}},T_{\text{mem}},T_{\text{others}})=\max(T_{\text{mem}},T_{\text{others}})
\end{equation}
where $T_{\text{mem}}$, $T_{\text{cmp}}$, $T_{\text{others}}$ and $T$ represent memory access, computation, other operation and total times, respectively. In this scenario, reducing computation time cannot influence total runtime.



\subsection{Fully Un-overlapped}\label{sec:anaunovl}
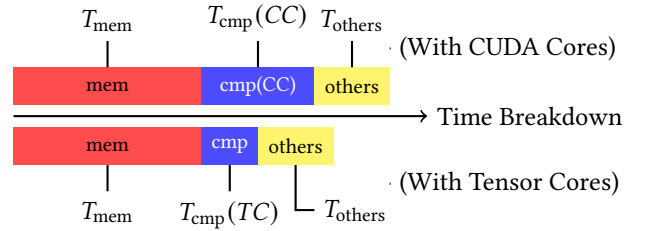
\begin{figure}[t!]
\vspace{-10pt}
\begin{tikzpicture}[scale=0.5, every node/.style={scale=1}]
    \draw[->, thick] (0,0) -- (11,0) node[right] {\textbf\footnotesize{Time Breakdown}};
    \fill[red!70] (0,0.3) rectangle (5,1.3) node[pos=.5,black] {\footnotesize{mem}};
    \fill[blue!70] (5,0.3) rectangle (8,1.3) node[pos=.5,white] {\footnotesize{cmp(CC)}};
    \fill[yellow!70] (8,0.3) rectangle (10,1.3) node[pos=.5,black] {\footnotesize{others}};

    \draw[thick] (2.5,1.3) -- (2.5,2) node[above] {$T_{\text{mem}}$};
    \draw[thick] (6.5,1.3) -- (6.5,2) node[above] {$T_{\text{cmp}}(CC)$};
    \draw[thick] (9,1.3) -- (9,2) node[above] {$T_{\text{others}}$};

    \draw[thick] (10,1.8) -- (10,1.8) node[right] {(With CUDA Cores)};

    \fill[red!70] (0,-0.3) rectangle (5,-1.3) node[pos=.5,black] {\footnotesize{mem}};
    \fill[blue!70] (5,-0.3) rectangle (6.5,-1.3) node[pos=.5,white] {\footnotesize{cmp}};
    \fill[yellow!70] (6.5,-0.3) rectangle (8.5,-1.3) node[pos=.5,black] {\footnotesize{others}};
    \draw[thick] (2.5,-1.3) -- (2.5,-2) node[below] {$T_{\text{mem}}$};
    \draw[thick] (5.75,-1.3) -- (5.75,-2) node[below] {$T_{\text{cmp}}(TC)$};
    \draw[thick] (7.5,-1.3) -- (7.5,-2.5) -- (8,-2.5) node[right] {$T_{\text{others}}$}; 

    \draw[thick] (10,-1.8) -- (10,-1.8) node[right] {(With Tensor Cores)};
\end{tikzpicture}
\vspace{-5pt}
\caption{Fully un-overlapped kernel time breakdown}\label{fig:breakout}
\end{figure}

For fully un-overlapped kernels (Figure~\ref{fig:breakout}):
\begin{equation}\footnotesize
T = T_{\text{cmp}}+T_{\text{mem}}+T_{\text{others}}
\end{equation}

With tensor cores providing speedup $\alpha$ ($\alpha=\tfrac{P(TC)}{P(CC)}$,$\alpha>1$): $T'_{\text{cmp}}(TC)=\tfrac{1}{\alpha}T_{\text{cmp}}(CC)$, yielding:
\begin{align}\footnotesize
Speedup&=\tfrac{T(CC)}{T(TC)} = \tfrac{T_{\text{cmp}(CC)}+T_{\text{mem}}+T_{\text{others}}}{\tfrac{1}{\alpha}T_{\text{cmp}(CC)}+T_{\text{mem}}+T_{\text{others}}} \\
&=1+\tfrac{{\alpha}-1}{1+{\alpha}\tfrac{T_{\text{mem}}+T_{\text{others}}}{T_{\text{cmp}}(CC)}} 
\\
&=1+\tfrac{{\alpha}-1}{1+{\alpha}\tfrac{T_{\text{cmp}}(CC)\times\tfrac{\mathbb{B}}{\mathbb{I}}+T_{\text{others}}}{T_{\text{cmp}}(CC)}} \\
&<1+\tfrac{\alpha-1}{1+\alpha(\tfrac{\mathbb{B}}{\mathbb{I}})}
\end{align}


\noindent\textbf{Tensor Core Upper Bound:} For memory-bound kernel, we have $T_{\text{cmp}}\to T_{\text{mem}}$:
\begin{align}\footnotesize
Speedup<1+\tfrac{\alpha-1}{1+\alpha}=2-\tfrac{2}{1+\alpha}
\end{align}
Example: FP64 Nvidia GPUs (with $\alpha=2$): $Speedup<1.33$.

\noindent Example: Assuming that $\alpha\to\infty$, we have $Speedup<2$.

\noindent\textbf{Workload Upper Bound:} we assume that $\alpha\to\infty$:
\begin{align}\footnotesize
Speedup <1+\tfrac{\mathbb{I}}{\mathbb{B}}
\end{align}
Example: $Speedup_{A100}(\text{GEMV})<1.05$.

\subsection{Summary}

Our analysis covers the two extremes of memory-computation overlap. Real-world kernels typically exhibit partial overlap, resulting in speedups between 1× and 1.33× for double precision. Performance differences beyond that would require memory access optimizations, which, we argue, function equally when applied to tensor and CUDA cores since both access data through the register file (As Figure~\ref{fig:tcarch} shows).




\section{Empirical Analysis}\label{sec:eval}

In this section, we use empirical analysis to verify our theoretical findings regarding tensor core performance on memory-bound kernels. We conduct experiments across multiple hardware platforms, detailed in Table~\ref{tab:env}, to systematically evaluate the performance relationship between tensor core and CUDA core implementations.

\begin{table}[t]
    \caption{Specifications of the experimental platforms.}
    \label{tab:env}
    \vspace{-14pt}
    \centering
    \footnotesize
    \begin{tabular}{|l|c|c|c|}
    \hline
    \multicolumn{2}{|c|}{\textbf{Metric}}             & \textbf{A100-80GB} & \textbf{GH200} \\\hline
    \multicolumn{2}{|c|}{\textbf{CUDA Version}}             & 12.1 & 12.6\\\hline
    \multicolumn{2}{|c|}{\textbf{L2 Cache (MB)}}        & 40               & 50           \\\hline
    \multicolumn{2}{|c|}{\textbf{Memory Bandwidth (TB/s)}}        & 1.94               & 4.00           \\\hline
    \multirow{2}{*}{\textbf{FP64 Peak (TFLOPS)}} &\textbf{CUDA Core}&  9.7    &  34.0  \\
                                            &\textbf{Tensor Core}& 19.5   &  67.0   \\\hline
    %
    \end{tabular}
    
\end{table}



\subsection{SCALE}

\begin{figure}[t!]
    \centering
    \includegraphics[width=\linewidth]{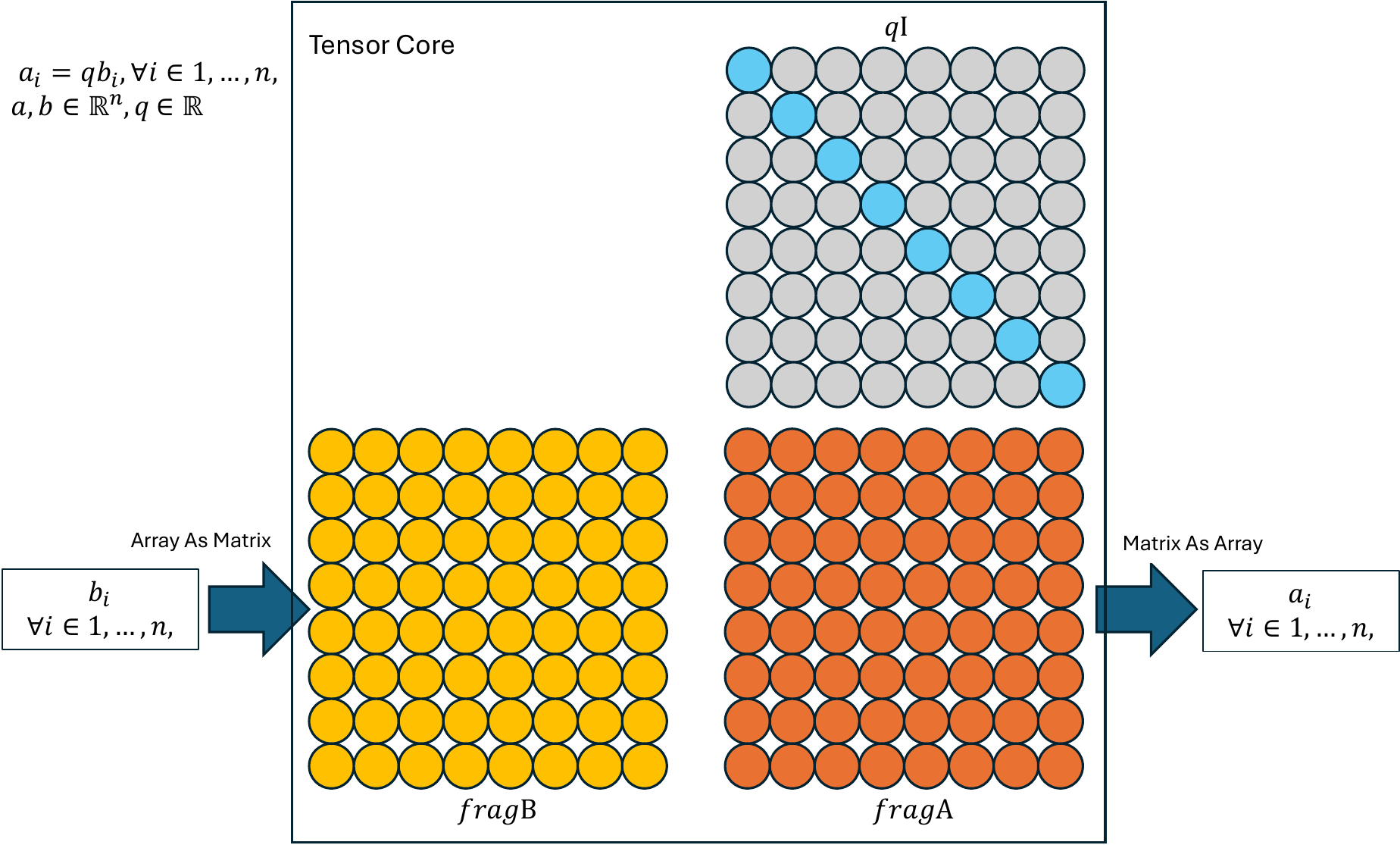}
    \vspace{-24pt}
    \caption{Tensor core SCALE implementation.}
    \label{fig:scaleimp}
\end{figure}

\noindent\textbf{Tensor Core Implementation:}
Inspired by work~\cite{navarro2020gpu}, we implement tensor core SCALE as matrix multiplication $A=B(qI)$, where $I$ is the identity matrix, as shown in Figure~\ref{fig:scaleimp}.
However, this implementation utilizes only $\tfrac{1}{max(m,n)}$ of the tensor core's compute capacity ($m$, $n$ = tensor core dimensions). For A100 and H100's $8\times4$ double precision tensor cores, only 1/8 of compute power is used:
$\mathbb{P}_{A100}(TC,SCALE)=2.4$, TFLOPS/s $\mathbb{P}_{GH200}(TC,SCALE)=8.37$ TFLOPS/s. This is lower than CUDA core performance. However, this should not significantly impact SCALE kernel performance, with or without overlap, given its extremely low operational intensity $\mathbb{I}$ (According to Section~\ref{sec:theory}).


\noindent\textbf{CUDA Core Implementation:}
The CUDA core baseline uses a ChatGPT-generated STREAM implementation{\footnote{\url{https://chatgpt.com/share/67570ae8-4554-8007-9f91-48f01722af85}}}, modified only to include warmup iterations.


\subsection{SpMV}

\begin{table*}[ht]
    \caption{Datasets for the SpMV benchmark (from DASP~\cite{10.1145/3581784.3607051}). The dataset is ranked by the non-zero value size}
    \vspace{-14pt}
\footnotesize
    {%
    \begin{tabular}[t]{|cccc|cccc|cccc|}
    \hline
        \textbf{Code} & \textbf{Name~\cite{davis2011university}} & \textbf{Rows} & \textbf{NNZ} \\ \hline
        \textbf{D1}   & dc2                                    & 116,835       & 766,396       \\
        \textbf{D2}   & scircuit                               & 170,998       & 958,936       \\
        \textbf{D3}   & mac\_econ\_fwd500                      & 206,500       & 1,273,389     \\
        \textbf{D4}   & conf5\_4-8x8-10                        & 49,152        & 1,916,928     \\
        \textbf{D5}   & mc2depi                                & 525,825       & 2,100,225     \\
        \textbf{D6}   & rma10                                  & 46,835        & 2,374,001     \\
        \textbf{D7}   & cop20k\_A                              & 121,192       & 2,624,331     \\ \hline
    \end{tabular}
    \begin{tabular}[t]{|cccc|cccc|cccc|}
    \hline
        \textbf{Code} & \textbf{Name~\cite{davis2011university}} & \textbf{Rows} & \textbf{NNZ} \\ \hline
        \textbf{D8}   & webbase-1M                             & 1,000,005     & 3,105,536     \\
        \textbf{D9}   & ASIC\_680k                             & 682,862       & 3,871,773     \\
        \textbf{D10}  & cant                                   & 62,451        & 4,007,383     \\
        \textbf{D11}  & pdb1HYS                                & 36,417        & 4,344,765     \\
        \textbf{D12}  & consph                                 & 83,334        & 6,010,480     \\
        \textbf{D13}  & shipsec1                               & 140,874       & 7,813,404     \\
        \textbf{D14}  & mip1                                   & 66,463        & 10,352,819    \\ \hline
    \end{tabular}
    \begin{tabular}[t]{|cccc|cccc|cccc|}
    \hline
        \textbf{Code} & \textbf{Name~\cite{davis2011university}} & \textbf{Rows} & \textbf{NNZ} \\ \hline
        \textbf{D15}  & pwtk                                   & 217,918       & 11,634,424    \\
        \textbf{D16}  & Si41Ge41H72                            & 185,639       & 15,011,265    \\
        \textbf{D17}  & in-2004                                & 1,382,908     & 16,917,053    \\
        \textbf{D18}  & Ga41As41H72                            & 268,096       & 18,488,476    \\
        \textbf{D19}  & eu-2005                                & 862,664       & 19,235,140    \\
        \textbf{D20}  & FullChip                               & 2,987,012     & 26,621,990    \\ 
        \textbf{D21}  & circuit5M                              & 5,558,326     & 59,524,291    \\ \hline
    \end{tabular}
    }
    \label{tab:matrxset}
\end{table*}

We evaluate performance using the same 21 representative sparse matrices from the DASP study~\cite{10.1145/3581784.3607051} (Table~\ref{tab:matrxset}).

\noindent\textbf{DASP~\cite{10.1145/3581784.3607051} (Tensor Core):} 
DASP, the SoTA tensor core SpMV implementation, employs a hybrid approach, categorizing matrix rows as long, middle, or small, and applies specialized processing strategies for each category, including row sorting for middle-length rows.

\noindent\textbf{cuSPARSE CSR~\cite{naumov2010cusparse} (CUDA Core):}
While formats with reordering (e.g., SELL-C-$\sigma$\cite{doi:10.1137/130930352}) might provide a more direct comparison to DASP, sorting can alter matrix characteristics and complicate performance analysis\cite{anzt2014implementing}. Therefore, we use the widely adopted cuSPARSE CSR format~\cite{naumov2010cusparse} as our baseline.

\subsection{Iterative Stencils}

\begin{table}[t]
    \vspace{-8pt}
    \centering
    \caption{\label{tab:stencilbench}Stencil benchmarks and domain sizes we use. A detailed description of the stencil benchmarks can be found in~\cite{zhao2019exploiting,rawat2016effective}.
    }
    \vspace{-10pt}
    {%
    \footnotesize
\begin{tabular}{l| c c c}
\toprule
\multirow{2}{*}{}&\multicolumn{3}{c}{\textbf{domain}(\textbf{temporal blocking depth})} \\
\cmidrule(r){2-4} 
        & ConvStencil~\cite{10.1145/3627535.3638476} & Brick~\cite{zhao2019exploiting} & EBISU~\cite{10.1145/3577193.3593716}  \\
\midrule
\textbf{2d5pt}   & $10240^2$(3) &  -       & $9000^2$(3)  \\
\textbf{2d13pt}  & $10240^2$(1)   &   -      &  $9000^2$(1)    \\
\textbf{2d9pt}   & $10240^2$(3) &    -     &  $9000^2$(3) \\
\textbf{2d49pt}  & $10240^2$(1)   &     -    &  $9000^2$(1)    \\
\textbf{3d7pt}   & $1024^3$(3)   & $512^3$(1)   & $234\times312\times2560$(3)     \\
\textbf{3d27pt}  & $1024^3$(3)  & $512^3$(1)   & $234\times312\times2560$(3)     \\
\bottomrule
    \end{tabular}

    }
    \label{tab:domain}
\end{table}

Stencil implementations were evaluated using ConvStencil~\cite{10.1145/3627535.3638476} benchmark suite (Table~\ref{tab:stencilbench}). 
Due to bugs in both ConvStencil's and LoRAStencil's Artifact Description/Artifact Evaluation (AD/AE) on the GH200 platform, our experiments are restricted to the A100 platform.

\noindent\textbf{ConvStencil~\cite{10.1145/3627535.3638476} (Tensor Core):}
ConvStencil leverages tensor cores by transforming stencil computation into matrix-matrix multiplication, incorporating temporal blocking through kernel fusion. We evaluate using their default configuration and domain sizes.

\noindent\textbf{LoRAStencil~\cite{lorastencil} (Tensor Core):} LoRAStencil applies Low-Rank adaptation to reduce stencil computational redundancy. While innovative, their artifact evaluation relies on assumptions about the rank of stencil weights, which limits its practical applicability. Due to these constraints, we use their published performance results for comparison.
%

\noindent\textbf{Brick~\cite{zhao2019exploiting} (CUDA Core):} Baseline CUDA Core implementation without temporal blocking. We evaluated it using the default configuration.

\noindent\textbf{EBISU~\cite{10.1145/3577193.3593716} (CUDA Core):} EBISU is a state-of-the-art CUDA Core implementation that incorporates temporal blocking. To ensure a fair comparison, we configured EBISU's temporal blocking parameters to match those of ConvStencil.


\begin{figure}[t!]
\vspace{-8pt}
    \centering
    \includegraphics[width=\linewidth]{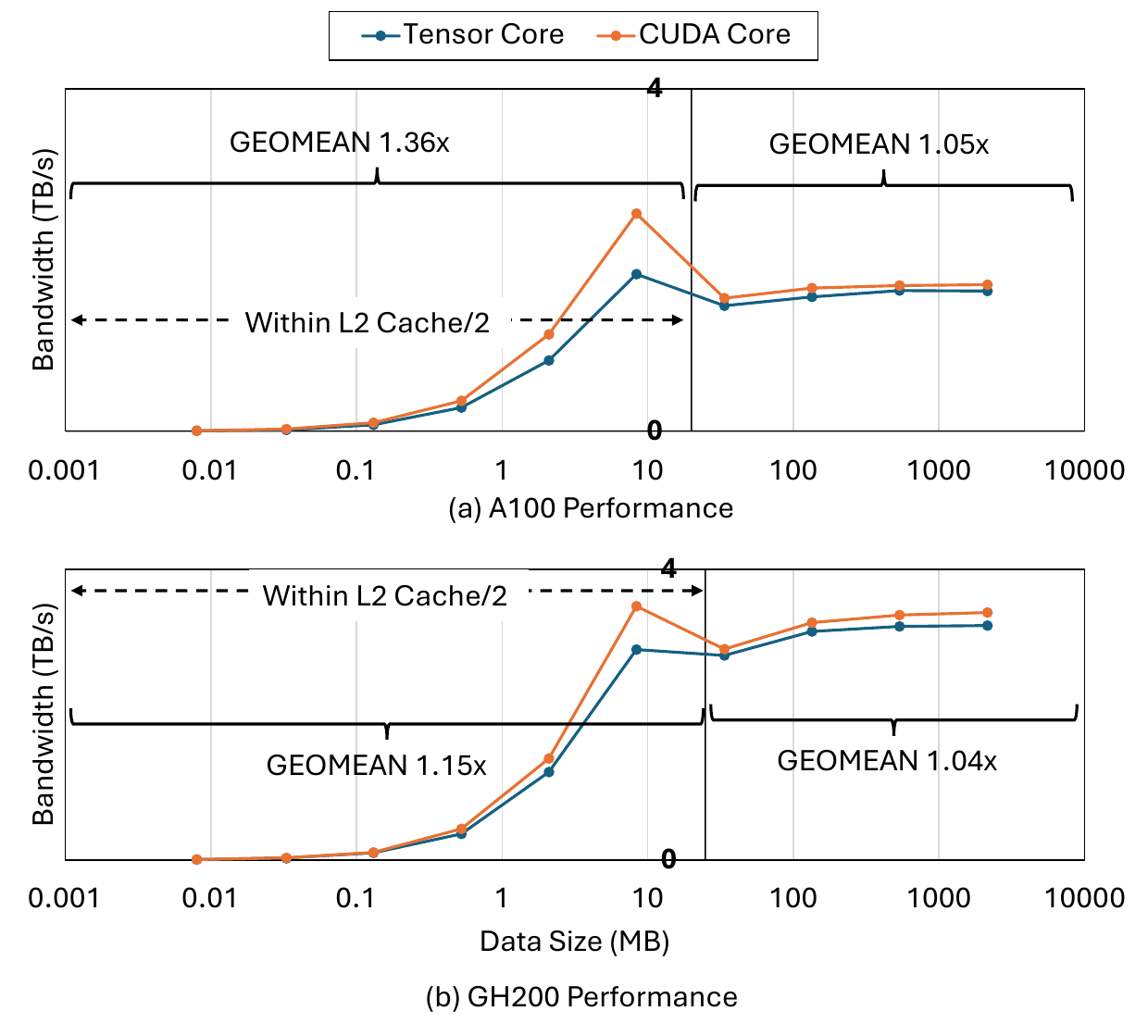}
    \vspace{-25pt}
    \caption{SCALE performance evaluation on A100 (top) and GH200 (bottom). We show the speedup geometric mean (GEOMEAN) of the CUDA cores over the tensor cores is reported for input domain sizes larger and smaller than half of the L2 cache, respectively.}
    \label{fig:scaleeval}
\end{figure}
\begin{figure*}[ht!]
\vspace{-15pt}
    \centering
    \includegraphics[width=\linewidth]{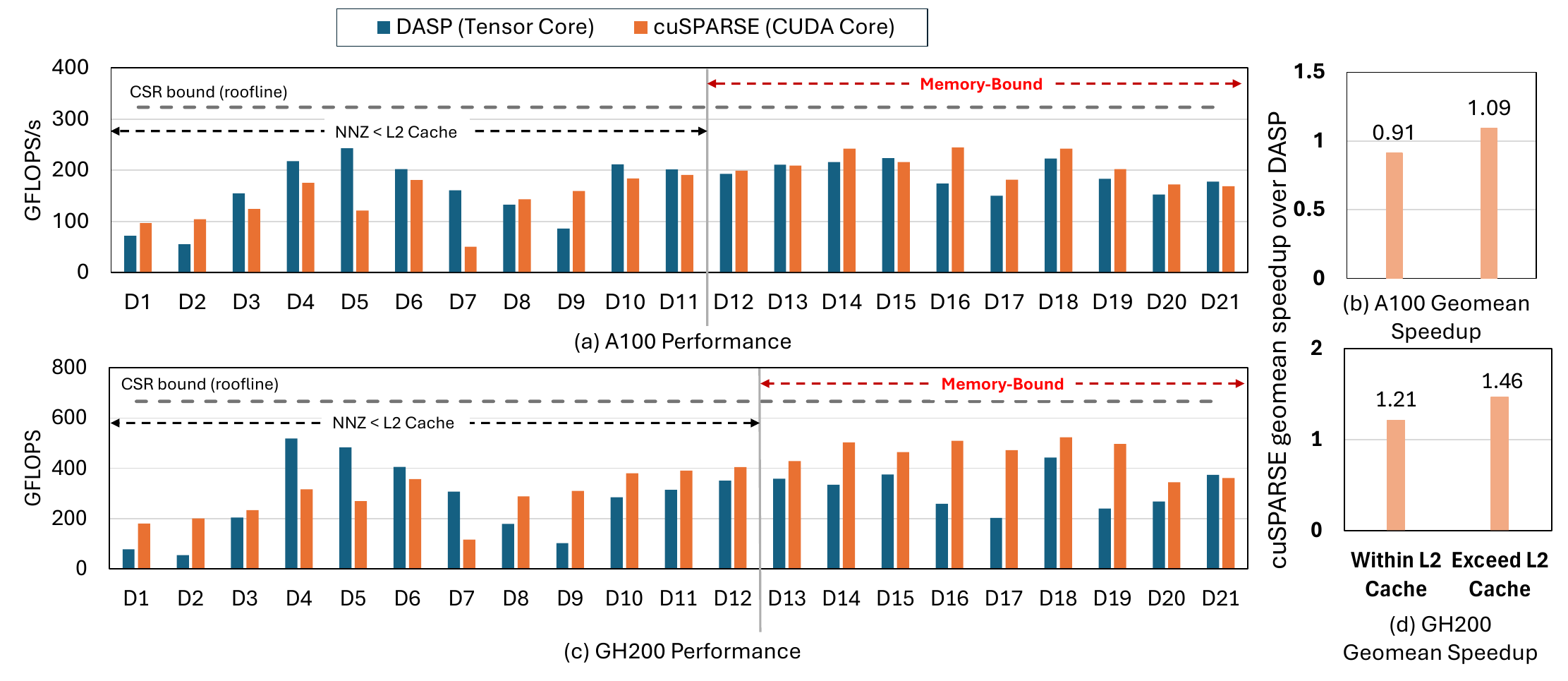}
    \vspace{-30pt}
    \caption{Comparison of cuSPARSE (CUDA Core) and DASP (Tensor Core) in sparse matrix-vector multiplication on A100 (top) and GH200 (bottom). (a) \& (c) on the left report the performance in effective flops; (b) \& (d) on the right report the geometric mean speedup of cuSPARSE over DASP.}
    \label{fig:spmv}
\end{figure*}
\begin{figure*}[ht!]
\vspace{-12pt}
    \centering
    \includegraphics[width=\linewidth]{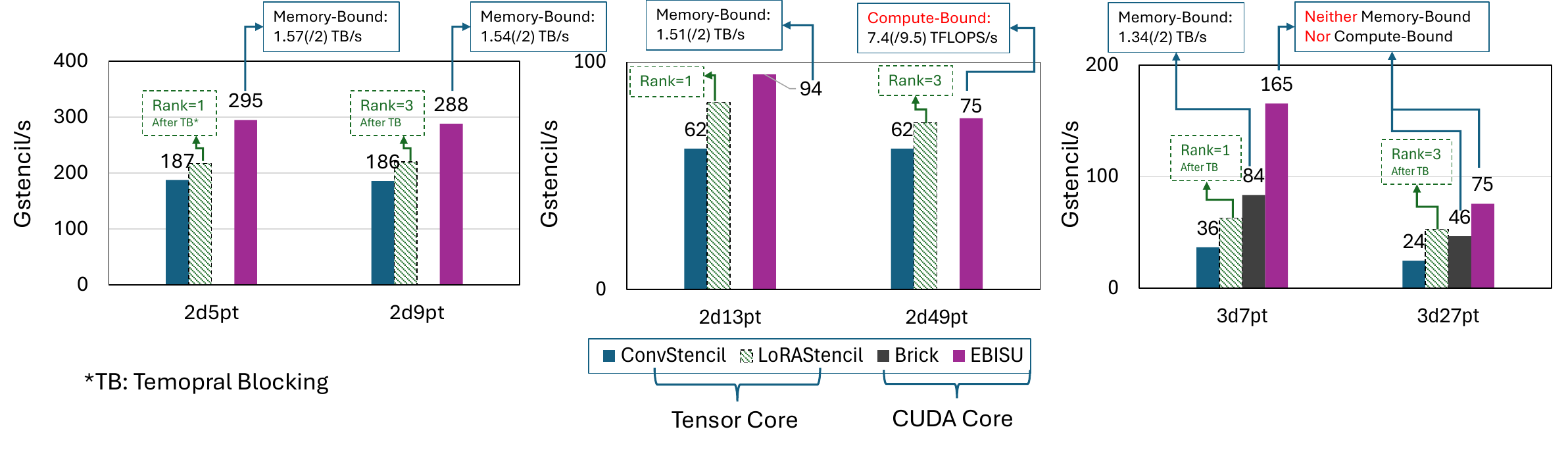}
    \vspace{-30pt}
    \caption{Comparison of EBISU and Brick and ConvStencil and LoraStencil on A100 using a suite of stencil benchmarks. For LoRAStencil~\cite{lorastencil}, performance data and their assumed rank values from their artifact evaluation are included.}
    \label{fig:stencilev}
\end{figure*}

\subsection{Evaluation}


Figures~\ref{fig:scaleeval}, \ref{fig:spmv}, and \ref{fig:stencilev} present performance comparisons for SCALE, SpMV, and stencil kernels respectively. 

\noindent\textbf{SCALE:} Figure~\ref{fig:scaleeval} reveals consistent, though modest, performance degradation when using tensor cores compared to CUDA cores. Given that the computational time difference is negligible, this performance gap likely arises from suboptimal memory access patterns associated with tensor core usage on current GPU architectures.

\noindent\textbf{SpMV:} Figure~\ref{fig:spmv} demonstrates that for datasets exceeding the L2 cache size, cuSPARSE (CUDA core) outperforms DASP (tensor core) on average.

\noindent\textbf{Stencils:} Figure~\ref{fig:stencilev} shows that equivalently optimized tensor core implementations generally underperform their CUDA core counterparts.

\noindent\textbf{Summary:} While our goal was to verify the theoretical bounds, empirical evaluation reveals that tensor core implementations usually underperform their CUDA core counterparts.



\subsection{Other Observations}
\noindent\textbf{L2 Cache Impact:} L2 cache interactions exhibit nontrivial between implementations. For SCALE, tensor core performance degradation intensifies within L2 cache bounds. Conversely, DASP demonstrates improved performance for cache-resident data, suggesting crucial performance implications of L2 cache optimization.

\noindent\textbf{Compute-Bound Cases:} The 2d49pt stencil, which is compute-bound on A100, shows comparable performance between tensor and CUDA core implementations. However, on GH200, the same kernel becomes memory-bound, where CUDA cores theoretically maintain superior performance.

\noindent\textbf{Resource-Constrained Cases:} 3D stencils and high-order 2D stencils typically encounter bottlenecks beyond memory or compute limitations, such as register capacity, cache capacity, or cache bandwidth~\cite{10.1145/3577193.3593716}. Since tensor core optimizations target only computational aspects, they provide no inherent advantage for these resource-constrained workloads. As expected, our evaluation shows no performance benefits from tensor core implementations in these stencil benchmarks.

\section{Key Takeaways}\label{sec:tackaway}
We highlight the key takeaways from a practical perspective: 

\begin{itemize}
    \item Identifying kernel characteristics (compute-bound/ memory-bound) (Section~\ref{sec:workload}). 
    \item Compute-bound
    \begin{itemize}
        \item Tensor cores remain advantageous for compute-bound operations.
    \end{itemize}
    \item Memory-bound
        \begin{itemize}
        \item Prioritizing CUDA cores for memory-bound kernels due to their simplicity and effectiveness (Section~\ref{sec:eval}).
        \item Focusing on memory access optimizations, e.g. cache-aware algorithms and reducing memory traffic~\cite{10.1145/3577193.3593716,10.1145/3577193.3593705}.
        \item Prioritizing pipeline and overlap optimizations before considering tensor core adoption (Section~\ref{sec:anaovl}). 
        \item Considering theoretical limits: tensor cores provide at most $1.33$× speedup in double precision, with a $2$× ceiling assuming infinite tensor core computational speedup (Section~\ref{sec:anaunovl}).
    \end{itemize}
\end{itemize}

\section{Conclusion}\label{sec:conclude}

Through systematic theoretical and empirical analysis, we demonstrate that leveraging the tensor cores for computation in memory-bound kernels fails to deliver sound performance benefits. 
Our theoretical analysis establishes an upper bound of $1.33$× speedup for double-precision memory-bound kernels, while empirical results across SCALE, SpMV, and stencil show that tensor core implementations usually underperform their CUDA core counterparts. 
While these findings may temper expectations for using tensor cores in memory-bound kernels, efforts leveraging tensor cores still provide valuable insights for the design and utilization of matrix processing units in broader contexts.


\begin{acks}
This material was supported by the U.S. Dept. of Energy, Office
of Science, Advanced Scientific Computing Research (ASCR), under contracts DE-AC02-06CH11357 and DE-SC002\newline 4207.
\end{acks}
\bibliographystyle{ACM-Reference-Format}
\bibliography{acmart}
\end{document}